\documentclass{article}
\usepackage{standalone}
\usepackage{amsmath,amssymb}
\usepackage{amsthm}
\usepackage{subcaption}
\usepackage{algorithm}
\usepackage{soul}
\usepackage{algpseudocode}
\UseRawInputEncoding
\algblock{ParFor}{EndParFor}
\algnewcommand\algorithmicparfor{\textbf{parfor}}
\algnewcommand\algorithmicpardo{\textbf{do}}
\algnewcommand\algorithmicendparfor{\textbf{end\ parfor}}
\algrenewtext{ParFor}[1]{\algorithmicparfor\ #1\ \algorithmicpardo}
\algrenewtext{EndParFor}{\algorithmicendparfor}
\usepackage{arxiv}
\usepackage[utf8]{inputenc} 
\usepackage[T1]{fontenc}    
\usepackage{hyperref}       
\usepackage{url}            
\usepackage{booktabs}       
\usepackage{amsfonts}       
\usepackage{nicefrac}       
\usepackage{microtype}      
\usepackage{lipsum}		
\usepackage{graphicx}
\usepackage{wasysym}

\newcommand*{\vpointer}{\hbox{\scalebox{1}{\Huge\pointer}}}
\usepackage[
    natbib=true,
    style=numeric,
    sorting=none
]{biblatex}
\hypersetup{
    colorlinks=true,
    linkcolor=blue,
    filecolor=blue,      
    urlcolor=blue
}
\newcommand{\be}{\begin{equation}}
\newcommand{\ee}{\end{equation}}
\newcommand{\bea}{\begin{eqnarray}}
\newcommand{\eea}{\end{eqnarray}}
\newcommand{\bean}{\begin{eqnarray*}}
\newcommand{\eean}{\end{eqnarray*}}

\title{Efficient Large-Scale Simulation of Fish Schooling Behavior Using Voronoi Tessellations and Fuzzy Clustering
}
\author{Salah Alrabeei\\
Department of Computer Science\\
Western Norway Univ. of Applied Sciences\\
Bergen, Norway\\
\texttt{salah.alrabeei@hvl.no}\\
\And
\href{https://orcid.org/0000-0003-0646-9830}{\includegraphics[scale=0.06]{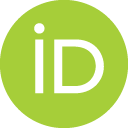}\hspace{1mm}Sam~Subbey}\thanks{Also at Department of Computer Science, Western Norway Univ. of Applied Sciences,	PO Box 7030, 5020 Bergen, Norway}\\
Inst. of Marine Research\\
PO Box 1870\\
5817 Bergen, Norway\\
\texttt{samuels@imr.no}\\
\And
Talal Rahman\\
Department of Computer Science\\
Western Norway Univ. of Applied Sciences\\
Bergen, Norway\\
\texttt{talal.rahman@hvl.no}\\
}

\hypersetup{
pdftitle={A template for the arxiv style},
pdfsubject={q-bio.NC, q-bio.QM},
pdfauthor={David S.~Hippocampus, Elias D.~Striatum},
pdfkeywords={First keyword, Second keyword, More},
}
\begin{document}
\maketitle
\begin{abstract}
This paper introduces an efficient approach to reduce the computational cost of simulating collective behaviors, such as fish schooling, using Individual-Based Models (IBMs). The proposed technique employs adaptive and dynamic load-balancing domain partitioning, which utilizes unsupervised machine-learning models to cluster a large number of simulated individuals into sub-schools based on their spatial-temporal locations. It also utilizes Voronoi tessellations to construct non-overlapping simulation subdomains. This approach minimizes agent-to-agent communication and balances the load both spatially and temporally, ultimately resulting in reduced computational complexity.

Experimental simulations demonstrate that this partitioning approach outperforms the standard regular grid-based domain decomposition, achieving a reduction in computational cost while maintaining spatial and temporal load balance. The approach presented in this paper has the potential to be applied to other collective behavior simulations requiring large-scale simulations with a substantial number of individuals.
\end{abstract}

\section{Introduction}
Collective behavior is a natural phenomenon observed in various dynamic complex living systems, where large groups of autonomous individuals participate based on limited local information transferred between each other through local interaction or their reaction to the surrounding environment. Such individual behaviors spread through the system, producing a collective pattern. Collective behaviors can be observed in biological and ecological systems such as schooling in fish \cite{huth1992simulation}, flocking in birds \cite{erneholm2011simulation}, herding mammals \cite{alglave2014herding}, and human crowding \cite{warren2018collective}. They are also observed in microorganisms such as cell populations \cite{maire2015molecular}, physics \cite{treado2021bridging}, and robotics \cite{schranz2020swarm}.

Agent-based modeling (ABM) is a powerful tool used to simulate and capture such emergent collective behaviors in complex systems \cite{egan2013design}. Continuum-based models, in which system variables evolve according to continuous functions over time, have limitations in modeling and simulating such collective behaviors and capturing emerging patterns \cite{solar2011high}. Simulating realistic collective behaviors often requires a large number of agents. However, running large-scale agent-based simulations has been a major challenge in ABM. This challenge is being addressed through the use of high-performance parallel computing and domain decomposition algorithms, which can be efficiently utilized to run such simulations. In particular, domain decomposition techniques have been commonly used by the research community in this field. Large-scale simulation workloads can be implemented by allocating each computing resource (processor, cluster, etc.) to a part of the simulation domain.

In domain decomposition-based simulation, selecting the proper partitioning mechanism for the simulated agent system is the main challenge. Two main approaches are used in the literature: grid-based and cluster-based \cite{vigueras2010comparative}. The grid-based approach consists of dividing the simulation domain into non-overlapping and uniform subdomains, each with a set of individuals residing temporarily in that subdomain. The cluster-based approach groups simulated individuals based on specific criteria and solves each subdomain and its simulated individuals or cluster separately in various modes, including sequential, parallel, or disturbed versions \cite{solar2012proximity}. Both approaches can be either static or dynamic, but the static method fails to maintain load balancing throughout the simulation evolution \cite{solar2011high,labba2015towards}.

Different types of partitioning have been implemented in grid-based domain decomposition, with most techniques tailored to the specific needs of a particular application area \cite{youseff2008parallel,cordasco2018distributed,araki2022dynamic}. The simplest type of domain decomposition is to divide the simulation domain into equal rectangular subdomains, but other techniques have been developed, such as regular micro-cells grouped into uniform subdomains, striped decomposition, block-striped decomposition, or even irregular grid cells \cite{zhang2009fast, solar2010high, zhou2004parallel, stijnman2003partitioning, egorova2019parallel}. However, using a partitioning method without an explicit load-balancing process can lead to increased computational time due to the extensive information exchange between agents at subdomain boundaries \cite{egorova2019parallel}. 

K-means is a widely used unsupervised and non-deterministic learning algorithm for clustering large datasets. It has been particularly useful in low-dimensional data and as a partitioning technique in agent-based simulations \cite{sun2008clustering,na2010research}. For instance, it has been adapted to partition agent-based crowd and vortex particle simulations \cite{wang2009cluster,marzouk2005k}. Several improved versions of k-means-based partitioning techniques have also been developed to efficiently distribute simulation workloads across computing resources \cite{von2018balanced,zhu2020improved,wu2021k}.

However, the k-means algorithm becomes less efficient when it needs to account for individual movement in the simulation model \cite{labba2015towards}. Furthermore, it does not consider cluster cardinalities, resulting in uneven cluster sizes when the dataset has varying densities, such as in particle interactions where aggregation and repulsion are present \cite{shen2014distributed}. Consequently, the simulation workload may become unbalanced, leading to expensive and inefficient large-scale simulators.

This study introduces a hybrid cluster- and region-based partitioning approach in a fish schooling simulation model. Our method is based on fuzzy (soft) clustering and Voronoi tessellations, ensuring spatial and temporal load balancing. The algorithm can be implemented sequentially or in parallel and can run on a local or remotely distributed computing resource. In Section \ref{sec2}, we introduce the biological fish schooling behavior model. In Section \ref{sec3}, we describe our simulation domain partitioning and population cluster algorithm. We present simulation results and a comparison between the static rectangular decomposition grid-based partitioning method and our proposed algorithm in Section \ref{sec4}. Finally, we conclude by discussing future work in Section \ref{con}.
\section{Description of fish schooling model}
\label{sec2}
Collective behaviors, such as fish schooling, bird flocking, and mammal crowding, have been widely studied using an agent-based modeling approach. This remains an active area of research in computational simulations. One of the earliest models for studying collective behavior is the Boid (or Reynolds) model \cite{reynolds1987flocks}. In this model, each individual aligns its movement direction with its neighbors, moves away from those neighbors who are too close to avoid collisions, and then flocks together. Subsequent extensions of the Boid model have incorporated additional features, such as obstacle avoidance \cite{erra2004massive,tran2020switching}, predator avoidance \cite{barksten2013extending,chang2019investigating}, leadership \cite{hartman2006autonomous,alaliyat2014optimisation}, field preference attraction \cite{chen2006genetic,dmytruk2021safe}, and multiple interaction zones \cite{barbaro2009discrete}. Other variations include the type of neighboring measure used \cite{ballerini2008interaction}.

\begin{figure}[!ht]
    \centering
    \includegraphics[scale=0.5]{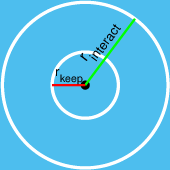}
  \caption{Regions of interaction, $r_{interact}$, and repulsion, $r_{keep}$, are centered on reference fish, $rf$ (black dot). \newline}
  \label{fig:regions}
\end{figure}

\begin{figure}[!ht]
    \begin{subfigure}[b]{0.32\textwidth}
 \centering   
    \includegraphics[scale=0.35]{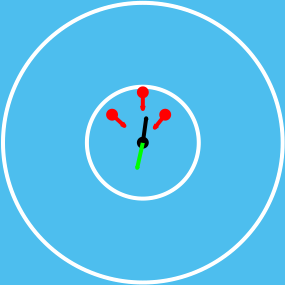}
  \caption{\small Direction change (black to green arrow) to avoid too close schoolmates (red dots).}
    \label{fig:collision}
  \end{subfigure}\hfill
  \begin{subfigure}[b]{0.32\textwidth}
   \centering
    \includegraphics[scale=0.35]{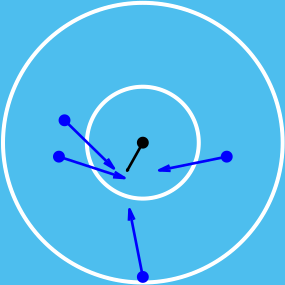}
  \caption{Movement towards the centroid of positions where schoolmates are located.}
  \label{fig:cenering}
  \end{subfigure}
    \hfill
    \begin{subfigure}[b]{0.32\textwidth}
    \centering
    \includegraphics[scale=0.35]{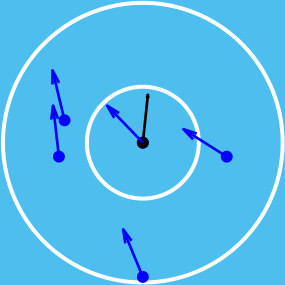}
  \caption{Direction change (black to blue arrow) to align movement with schoolmates.}
  \label{fig:attraction}
  \end{subfigure}
    \label{fig:interaction}
    \caption{Fish local interaction rules -- the reference individual fish $rf$  (black dot) whose original direction (black arrow) is influenced by neighboring fish (red and blue dots) whose movement directions are indicated by red and blue arrows.}
\end{figure}

Here, we adopt the Boid model \cite{reynolds1987flocks} to capture the collective behaviors of fish. Within a local range, each fish follows several interacting rules as illustrated in Figure \ref{fig:regions}. The cohesion rule is composed of two parts: (i) collision avoidance, which generates a repulsive force to ensure a minimum distance between schoolmates (see Figure \ref{fig:collision}), and (ii) school attraction, which produces a positive force to maintain school formation within the region of interaction (see Figure \ref{fig:attraction}). By combining these two rules, the fish school within a mutual range, achieving a cohesive group behavior. The cohesion force vector is denoted by $F_{u_{c}}$, and its mathematical expression is given in Eq.(\ref{eq:cohs}).

\begin{equation}
F_{u_{c}}= \frac{u_{c}}{\left\Vert u_{c}\right\Vert}\cdot\begin{cases} \frac{\left\Vert u_{c}\right\Vert - \text{r}_{\mathrm{keep}}}{\text{r}_{\mathrm{keep}}} &\text{if}\ \left\Vert u_{c}\right\Vert\leq \text{r}_{\mathrm{interact}},\\  0  &\text{Elsewhere}.\\  \end{cases}
\label{eq:cohs}
\end{equation}
 where $u_c$ is the vector from the reference individual to the nearest schoolmate, $\text{r}_{\mathrm{keep}}$ is the minimum safe distance to avoid collision with a schoolmate, and $\text{r}_{\mathrm{interact}}$ is the maximum distance to interact (range of the region of interaction).

The alignment rule in the fish school model is described by the force vector $F_{u_{a}}$ which represents the tendency of an individual fish to align its swimming velocity with its nearby schoolmates. The mathematical formula for $F_{u_{a}}$ is given by Eq.(\ref{eq:ali}):
\begin{equation}
F_{u_{a}} = \frac{1}{n} \sum_{i \in R_n} u_i - u_{rf},
\label{eq:ali}
\end{equation}
where $u_i$ is the velocity vector of individual $i$, $R_n$ is the set of $n$ individuals located within the interaction region of the reference individual $rf$, and $u_{rf}$ is the velocity vector of the reference individual.

In addition to the cohesion and alignment rules, there are also attraction and repulsion rules that are based on stimuli and risks in the environment, respectively \cite{chen2006genetic}. The force vectors $F_{u_{f}}$ and $F_{u_{r}}$ for attraction and repulsion, respectively, are given by Eq.(\ref{eq:food}) and Eq.(\ref{eq:risk}), where $u_f$ is the vector from the reference individual to the nearest stimuli source, and $u_r$ is the vector from the reference individual to the nearest risk source.
 \begin{eqnarray}
 \label{eq:food}
   F_{u_{f}} &=&  \frac{u_{f}}{\left\Vert u_{f}\right\Vert},\\
\label{eq:risk}     
F_{u_{r}} &=& \frac{u_{r}}{\left\Vert u_{r}\right\Vert}\cdot\begin{cases} \frac{\left\Vert u_{r}\right\Vert - \text{r}_{\mathrm{interact}}}{\text{r}_{\mathrm{interact}}} &\text{if}\ \left\Vert u_{r}\right\Vert\leq \text{r}_{\mathrm{interact}},\\  0  &\text{Elsewhere}.\\  \end{cases}
\end{eqnarray}
The ultimate strategy movement vector $V_{rf}$ of the $rf$ individual fish is defined as a weighted combination of all the influencing forces, which is expressed by Eq.~\ref{eq:finalF}, 
\begin{eqnarray}
\label{eq:finalF}
V_{rf}&=&\omega_{1}F_{u_{c}}+\omega_{2}F_{u_{a}}+\omega_{3} F_{u_{f}} + +\omega_{4} F_{u_{r}} 
\end{eqnarray}
where $\left\{\omega_{i}\right\}_{i=1}^4$, are the controlling weights that are based on the Spatiotemporal state of the individual surroundings. 
\section{Domain Decomposition Interaction Simulation Model}
\label{sec3}
Simulating collective behavior can be an extremely time-consuming process, often taking days, weeks, or even months. To reduce computational time, one effective solution is to partition the simulation domain into sub-domains. While domain decomposition may sound simple, its implementation requires careful consideration of issues that arise at inter-regional boundaries \cite{hanebutte1995traffic}.

In this work, we introduce an adaptive and load-balancing domain decomposition technique to efficiently simulate fish schooling behavior. Our approach utilizes an unsupervised machine learning algorithm to cluster the simulated fish into sub-schools based on their spatial-temporal locations. We then employ Voronoi tessellations to create bounded simulation subdomains that house the clustered population separately.

To mitigate communication overload between agents on the boundaries of subdomains and their neighboring subdomains, we replace subdomain-to-subdomain communication with area-to-area communication. This involves dividing each subdomain (polytope) into smaller areas based on their vertices, allowing agents within each area to communicate and interact solely with those in neighboring areas, rather than with the entire subdomain.

Our implementation approach incorporates several innovative ideas for leveraging this new algorithm to reduce complexity and enhance the efficient utilization of computational resources.

\subsection{Fuzzy c-Means Clustering}
\begin{figure}[ht]
    \centering
    \includegraphics[scale=0.4]{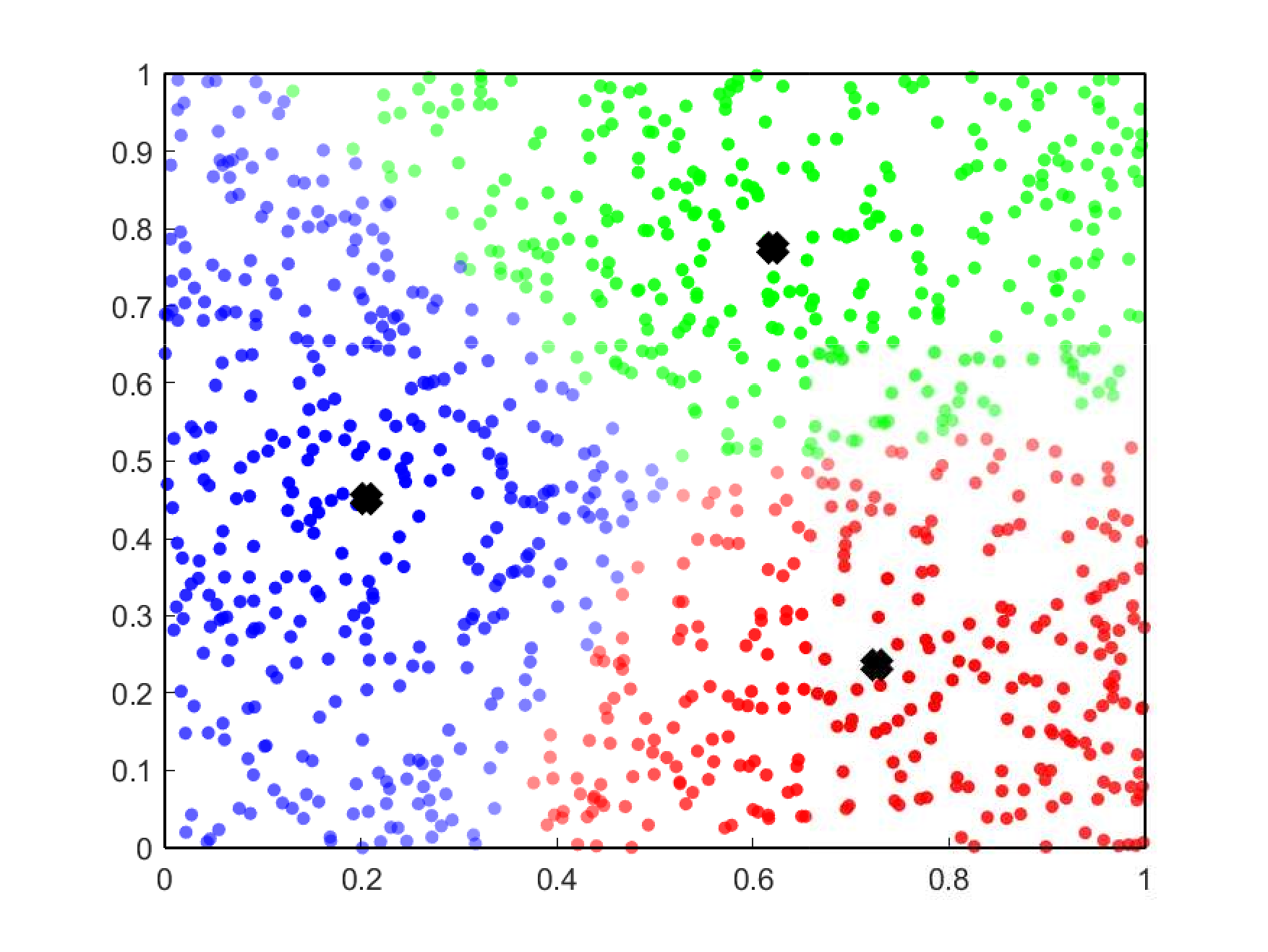}
    \caption{In this FCM clustering example, each cluster is centered at the black cross, with members of the same color (red, blue, or green). The intensity of the cluster color represents the association probability of members with the cluster. Members closer to the center have a higher association probability, while those near the boundary have a lower probability.
}
    \label{fig:FCM}
\end{figure}

Fuzzy c-Means (FCM) is a very popular soft (overlapping) clustering technique. FCM \cite{yang2019resource} is a clustering technique allowing each data point to belong to more than one cluster with association probabilities between 0 and
1 (see \textbf{Fig.} \ref{fig:FCM}).

The objective is to cluster a set of $N$ data point locations, denoted as $P = \{p_1, x_2, \ldots, x_N\}$, into $k$ clusters with centroids $C=\{c_1, c_2, \ldots, c_k\}$. The aim is to minimize the distances between cluster members while maximizing the distances between clusters. In other words, FCM formulates this as a minimization problem with the following objective function:

\begin{equation}
    J_m = \sum_{i=1}^N \sum_{j=1}^k \mu_{ij}^m \| p_i - c_j \|^2,  
\end{equation}
where $m>1$ is the fuzzy partition matrix exponent, which controls the degree of fuzzy overlap, and $\mu_{ij}$ is the probability of associating the point $x_i$ with the $j^{\mbox{th}}$ cluster. 

The FCM algorithm \cite{bezdek2013pattern} works by initially generating random association probabilities $\mu_{ij}$ of each point, $x_i$, and each cluster, $c_j$. The next step involves calculating the centroid of each data partition group (cluster) as in Eq. (\ref{eq:clustering}), 
\begin{equation}
    c_j = \frac{\sum_{i=1}^N \mu_{ij}^m x_i}{\sum_{i=1}^N \mu_{ij}^m},
    \label{eq:clustering}  
\end{equation}
followed by an update of the association probabilities according to Eq. ~(\ref{eq:prob}), 
\begin{equation}
    \label{eq:prob}
    \mu_{ij} = \frac{1}{    \sum_{r=1} ^{k} \bigg( \frac{\| x_i - c_j \|}{\| x_i - c_r \|} \bigg)^ \frac{2}{m-1}},
\end{equation}
and calculation of the object function, $J_m$. This iterative process continues until a stopping criterion is met, with the objective function $J_m$ consistently improving. In summary, the FCM algorithm proceeds as follows: during each iteration, it calculates association probabilities and cluster centers as per Eq. (\ref{eq:clustering}) and Eq. (\ref{eq:prob}), respectively. If the objective function $J_m$ falls below the specified threshold, the process concludes. Otherwise, the algorithm continues to update $\mu_{i,j}$, $c_i$, and $J_m$ until the stopping criterion is satisfied.

\subsection{Voronoi Diagram}
\begin{figure}[htbp]
    \centering
    \begin{tabular}{ccc}
        \includegraphics[scale=0.35]{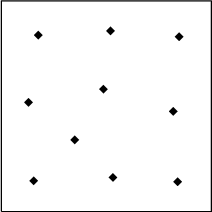} & $\vpointer$ & \includegraphics[scale=0.26]{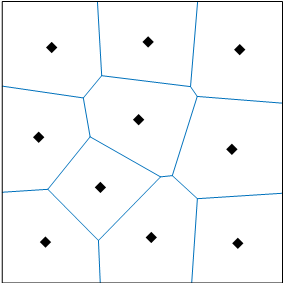} \\
        Random points & & Voronoi diagram
    \end{tabular}
    \caption{Illustration of Voronoi tessellation}
    \label{fig:voronoi}
\end{figure}

Voronoi Diagram (VD) \cite{boots1999spatial,pokojski2018voronoi} is defined as a set of Voronoi cells $ VD = \bigg \{ V(p_1),V(p_2),..., V(p_n) \bigg \}$, generated by the points $p_1,p_2,...,p_n$, where, given $d$ as the Euclidean distance,\\
$V(p_i) = \bigg\{ x |\; d(x,p_i) \le d(x,p_j) \;\; \text{for} j\neq i \bigg \}$  \\
Voronoi cells are non-overlapping, space-filling and the size (volume, area) of each cell is inversely proportional to the density of the generating points (see \textbf{Fig.}~\ref{fig:voronoi}). These properties of Voronoi cells are attractive in high-dimensional computations \cite{subbey2003strategy}. 

\subsection{Model Architecture}
Our model architecture comprises three key classes: \textbf{Agent}, \textbf{Region}, and \textbf{Domain}. 
\begin{itemize}
\item The Agent class represents the smallest unit in the simulation and includes attributes related to its spatial-temporal state.
\item The Region class defines a bounded area within the simulation domain that encompasses agents within its boundaries and neighboring regions. It also provides functions for managing communication between agents within its range and those in neighboring regions.
\item The Domain class represents a larger section of the simulation composed of multiple non-overlapping regions. It oversees the allocation of agents to regions and manages communication between these regions.
\end{itemize}

\subsubsection*{Region class}
\begin{figure}[htbp]
    \centering
    \begin{tabular}{ccccc}
        \includegraphics[scale=0.35]{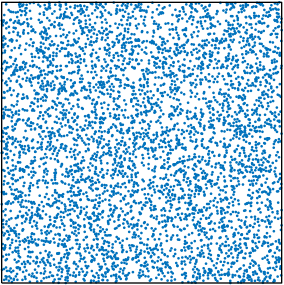} & $\vpointer$ & \includegraphics[scale=0.35]{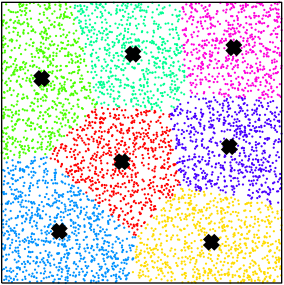} & $\vpointer$ & \includegraphics[scale=0.35]{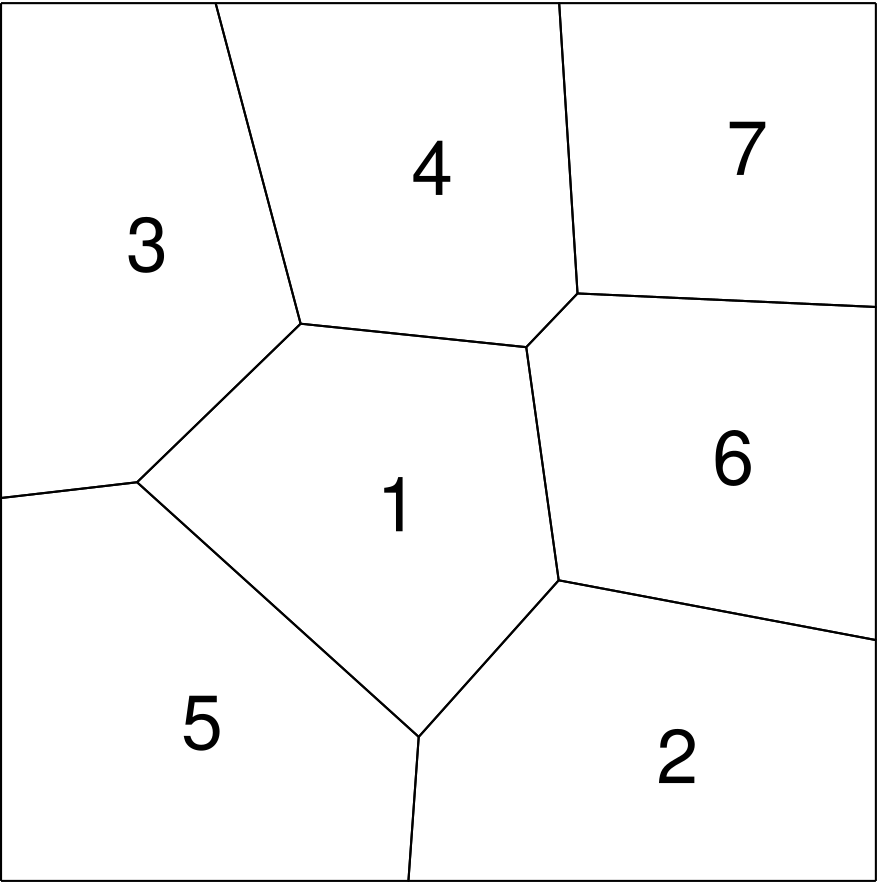}\\
        A large group agents & & Clustering the agents & & Partitioning the domain
    \end{tabular}
    \caption{Illustration of the CP algorithm}
    \label{fig:Pffft}
\end{figure}

In our simulation, we adopt an adaptive approach to region architecture, as opposed to a predefined "divide and conquer" method. The structure of regions is determined by the spatial-temporal distribution of agents. Our design for the region class aims to host an approximately equal number of agents, mitigating communication overhead in some regions while preventing others from remaining idle and causing redundant computation. This region construction process occurs in three steps: clustering, balancing, and partitioning.
\begin{enumerate}
    \item Clustering: We utilize the fuzzy c-Means (FCM) algorithm to group all fish agents into sub-schools, which occupy temporally non-overlapping regions within the simulation domain.
    \item Balancing: We ensure that each cluster has a balanced cardinality, maintaining an even distribution of agents.
    \item Partitioning: Finally, we employ the centroids of these balanced clusters to create non-overlapping Voronoi bounded regions that encompass the clustered agents (as shown in Fig. \ref{fig:Pffft}).
\end{enumerate}
\subsubsection*{Communication between agents}
Effective communication between agents to share information about their state and the surrounding environment is crucial in individual-based modeling. However, it can also significantly contribute to computational costs. Our CP algorithm and the internal architecture of our region class enable us to minimize redundant agent-to-agent communication.

\begin{figure}[!ht]
    \centering
    \includegraphics[scale=0.3]{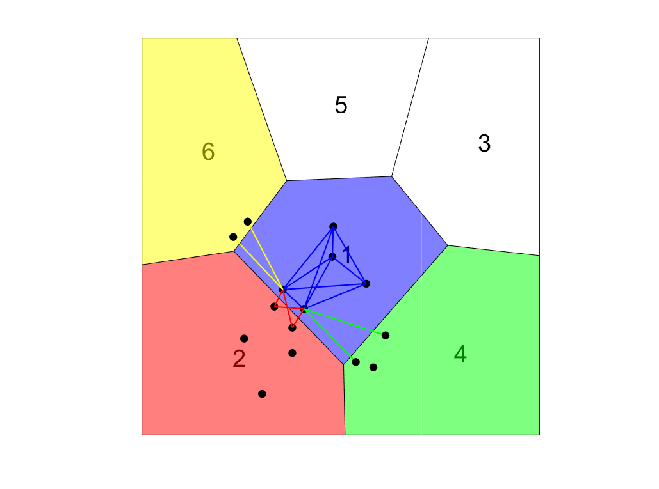}
    \caption{Agent-to-agent communication occurs at two levels within the region. Internal Communication: Agents within the same region (light blue) exclusively communicate with one another (blue communication graph); External Communication: External agents extend their communication to agents in neighboring regions who share the same vertices.}
    \label{fig:communcation}
\end{figure}

We implement two scales of communication per simulation time step:
\begin{enumerate}
    \item Region Scale (School-Scale): Each school within a region informs other schools in the entire fish population about member agents entering or leaving. Our adaptive region construction and communication are performed simultaneously using the CP algorithm.
    \item Individual Scale: Agents within each region communicate with those in their neighboring regions. Based on our region class's internal architecture, each agent only communicates with agents in the same region and those assigned to the nearest vertex of the neighboring region. This optimized communication strategy is illustrated in Fig. \ref{fig:communcation}.
\end{enumerate}
\subsubsection*{Assign Agent to Region}
Our Cluster-Partitioning (CP) algorithm offers several advantages beyond adaptive region definition. It not only partitions agents into regions but also identifies inter-region neighbors during the clustering step, eliminating the need for additional assignment and organization steps.

To minimize communication overhead between agents at region boundaries, we implement vertex-based partitioning within each region (polytope). This reduces communication from region-to-region to area-to-area. The classification of agents as internal or external is based on their interaction range. Agents with an interaction range confined within their region are classified as internal, while those reaching neighboring regions are classified as external (refer to Fig. \ref{fig:internal_external_agents}).

This classification is determined by constructing a triangle using the agent's location and the nearest two vertices of the region's corners. We then check whether the height of the triangle exceeds the interaction radius (as shown in Fig. \ref{fig:internal_external_agents_formula}).

\begin{minipage}{0.4\textwidth}
\begin{figure}[H]
\centering
\includegraphics[scale=0.34]{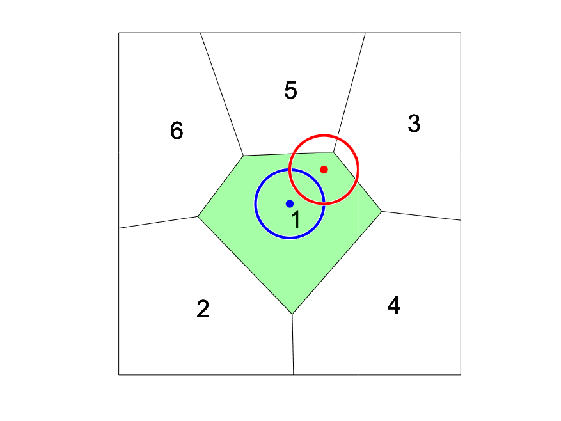}
\caption{Internal agents are represented by blue filled circles, while external agents are denoted by red filled circles. The interaction radii of internal, and external agents are indicated by blue and red circles, respectively.
}
\label{fig:internal_external_agents}
\end{figure}
\end{minipage}
\begin{minipage}{0.2\textwidth}
\hspace{0.01cm}
\end{minipage}
\begin{minipage}{0.4\textwidth}
\begin{figure}[H]
\centering
\includegraphics[scale=0.25]{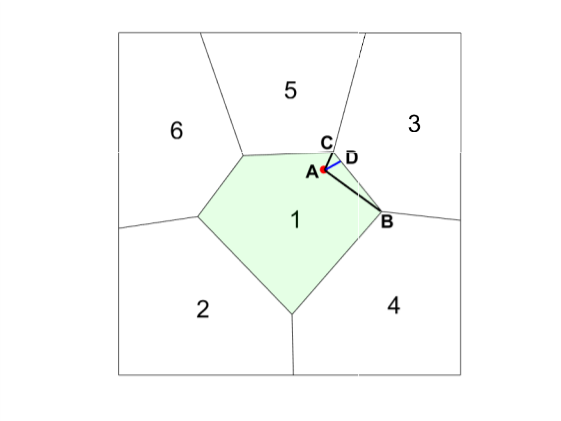}
\caption{The vertices of triangle ABC are such that A represents the agent's location, while B and C correspond to the agent's nearest two region corners. The classification of an agent as internal or external, is based on the triangle's height, $\overline{AD}$.}
\label{fig:internal_external_agents_formula}
\end{figure}
\end{minipage}

Equation~(\ref{eq:classifer1}) is mathematical and functional representation of a classifier: 
\begin{equation}
     f(\textit{Area},b) =
    \begin{cases}
      Internal & \text{if }  \hspace{4dd}  \text{\textit{Area}}/b\le R\\
      External & \text{if}  \hspace{4dd}  \text{\textit{Area}}/b > R \\
    \end{cases}   
    \label{eq:classifer1}
\end{equation}
where {\textit{Area}} and \textit{b} are the area and base of the triangle \textit{ABC}, and \textit{R} is the radius of interaction.  

\begin{figure}[!ht]
    \centering
    \includegraphics[scale=0.4]{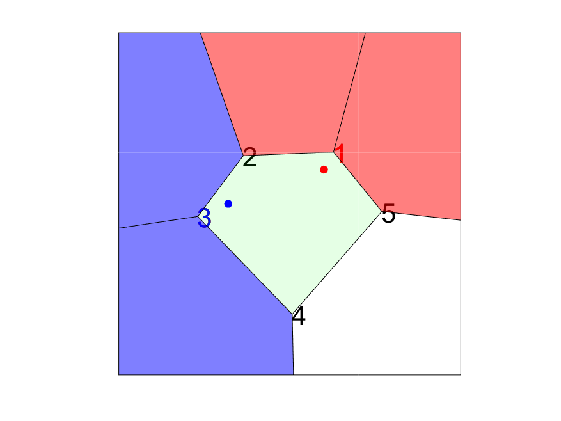}
    \caption{Two agents, represented by red and blue filled circles, are classified and assigned based on their proximity to region vertices (1 for the red agent and 3 for the blue agent). These external agents have interaction ranges extending to neighboring regions, indicated by the light red and purple regions, respectively}
    \label{fig:2nd_classifier}
\end{figure}

The second classifier categorizes each external agent based on its nearest region corner. This classification serves to eliminate redundant communication overhead from agents in distant neighboring regions, ensuring that agents communicate only with the essential regions (refer to Fig. \ref{fig:2nd_classifier}).

\section{Simulation experiments}
\label{sec4}
Domain decomposition-based simulation has the potential to significantly reduce computational costs, and researchers in this field have devoted substantial efforts to identifying efficient techniques for adaptive workload redistribution throughout the simulation. Workload can be quantified based on the number of agents in each subdomain or the computational time spent on each subdomain, as discussed in \cite{egorova2019parallel}.

In our simulations, we measured workload in three ways: across the entire simulation's duration to assess overall temporal load balancing, within each subdomain to evaluate spatial load balancing, and at each simulation time step for temporal load balancing.

Our partitioning algorithm demonstrates superior performance compared to regular grid-based partitioning. It excels in terms of total running time and the preservation of both spatial and temporal load balancing. Notably, our algorithm is versatile and suitable for parallel simulation or distributed cloud computing, although we exclusively utilized sequential implementations in our simulations.

All algorithms and simulations were executed on the MATLAB 2020b platform, operating on a PC equipped with an Intel Core i7, 2.6 GHz CPU, and 16 GB RAM.

\begin{figure}[!ht]
    \begin{subfigure}[b]{0.3\textwidth}
 \centering   
    \includegraphics[scale=0.21]{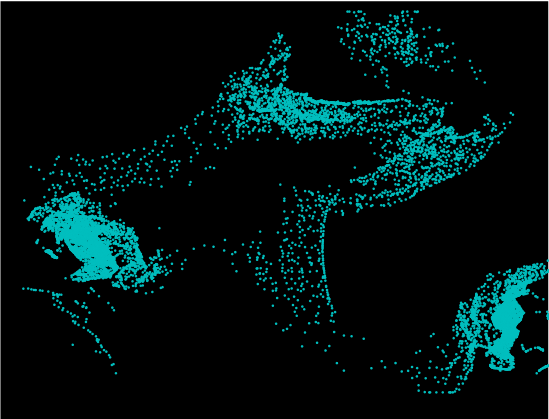}
  \caption{Spatial distribution of simulated fish}
  \end{subfigure}\hspace{0.2cm}
  \begin{subfigure}[b]{0.36\textwidth}
   \centering
    \includegraphics[scale=0.25]{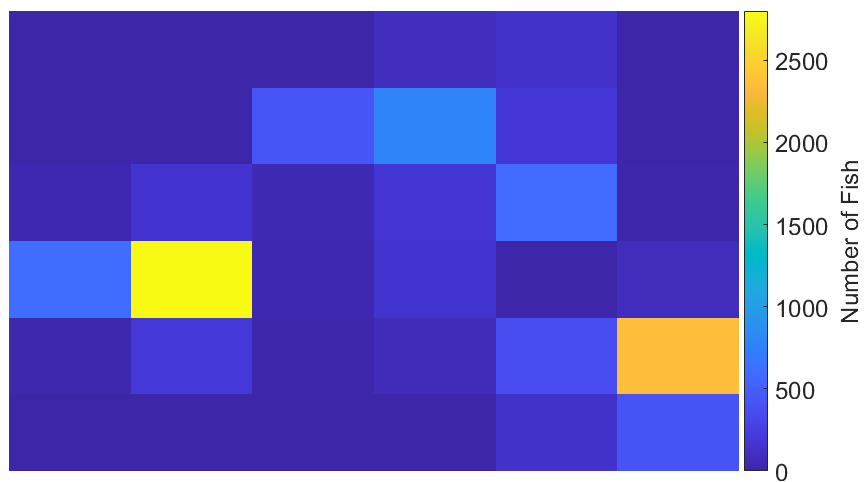}
  \caption{Rectangle-based partitioning}
  \end{subfigure}
   \hspace{0.2cm}
    \begin{subfigure}[b]{0.3\textwidth}
    \centering
    \includegraphics[scale=0.22]{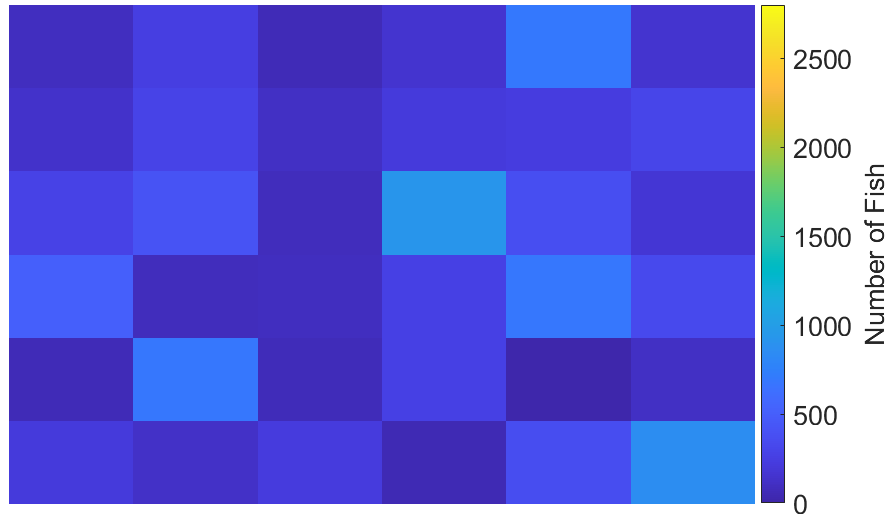}
  \caption{Cluster-based partitioning}
  \end{subfigure}
    \caption{Illustration of how ten thousand agents (in a) can be distributed over 36 subdomains constructed based on grid-based (in b) or cluster-based (in c)  partitioning.}
     \label{fig:com_heatmap}
\end{figure}

In contrast to the static rectangular grid-based approach \cite{vigueras2010comparative}, our proposed method ensures spatial and temporal load balancing, regardless of the initial agent distribution and simulation duration (as depicted in Fig. \ref{fig:com_heatmap}).
For our initial simulation experiment, we assessed the computational performance of three implementation methods: no partitioning, regular grid-based partitioning (square), and partitioning using our CP algorithm. We conducted four simulations for each method, each involving a different number of simulated fish.

Table \ref{Tab:com} illustrates the results, demonstrating that when the number of simulated fish is small, partitioning is unnecessary. However, as the number of simulated fish increases, the non-partitioning method becomes significantly more time-consuming than the partitioned methods. This disparity arises because the communication between simulated fish exhibits an order of $O(N^2)$, indicating that with a higher number of fish, more time is required for communication.
\begin{minipage}{0.4\textwidth}
\begin{table}[H]
\centering
\begin{tabular}{|c|c|c|c|}
\hline
 N/Type & NDD & RDD & VDD    \\ \hline
1e+02 & 0.17 & 0.27 & 2.4  \\ \hline
1e+03 & 13.08 & 7.85 
 & 6.60 \\ \hline
1e+04 & 1.4 e3 &  718.38 & 258.48  \\ \hline
2e+04 & 5.51 e+03 & 2.14e+03
& 1.02e+03 \\ \hline
\end{tabular}
\caption{
Computational CPU time costs, measured in seconds, for fish simulations employing various implementation methods. These simulations involve different numbers of simulated fish. Specifically, 'NDD' represents no domain decomposition, 'RDD' stands for regular domain decomposition, and 'VDD' denotes Voronoi domain decomposition, facilitated by the CP algorithm.
}
\label{Tab:com}
\end{table}
\end{minipage}
\begin{minipage}{0.05\textwidth}
\hspace{0.1cm}
\end{minipage}
\begin{minipage}{0.47\textwidth}
 \begin{figure}[H]
     \centering
     \includegraphics[scale=0.45]{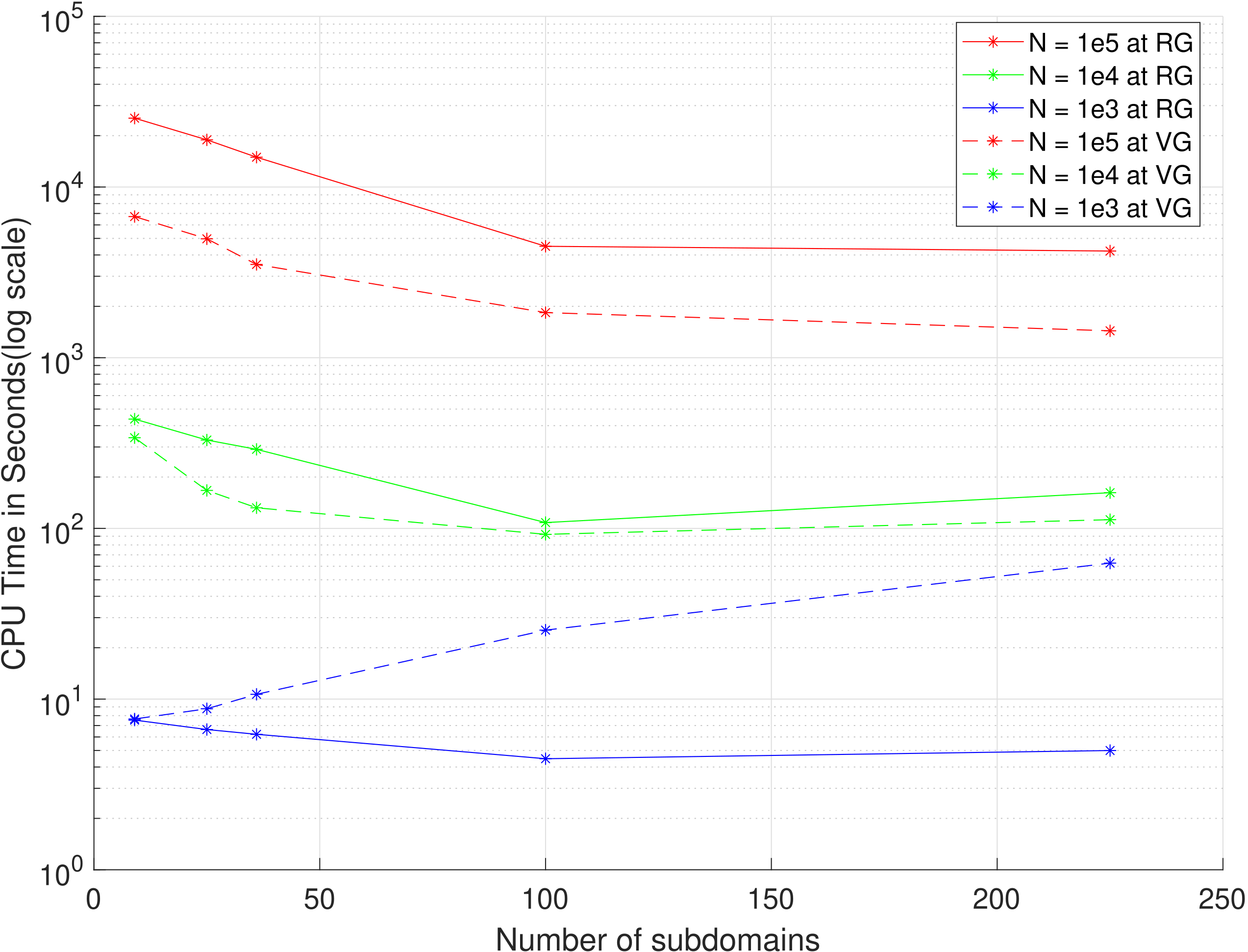}
     \caption{A computation cost comparison between regular grids (RDD)  and the Voronoi irregular grids (VDD) domain decomposition implementation. The figures show the CPU time as a function of the number of subdomains and the size of the simulated fish school}
     \label{fig:RG_VG_comp}
 \end{figure}
\end{minipage}

\subsection{Load balancing performance comparison: Regular Cells vs. Voronoi Cells}
\begin{figure}[H]
    \centering
    \includegraphics[scale=0.6]{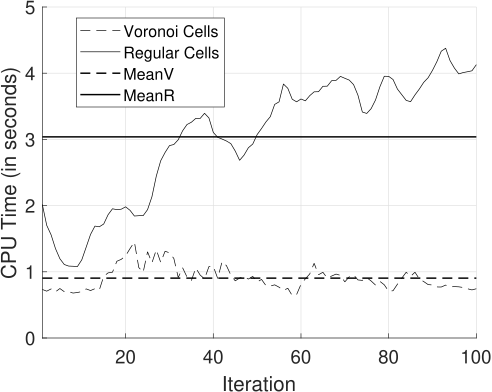}
    \caption{The computation cost profile of the fish simulation is depicted for both RDD (solid lines) and VDD (dashed lines) implementations. The stippled lines represent the CPU time for 100 time steps, while the bold lines denote the average CPU time.}
    \label{fig:CPU time avg}
\end{figure}

We conducted a load balancing performance comparison between our Voronoi-based domain decomposition (VDD) and the regular grid-based domain decomposition (RDD) methods. As depicted in Fig. \ref{fig:CPU time avg}, the VDD implementation excels in achieving a well-balanced workload distribution among subdomains, resulting in a lower variance of CPU time across subdomains when compared to the RDD implementation. This advantage is primarily attributed to VDD's adaptability in adjusting subdomain size and shape based on the spatial distribution of fish, which promotes an even workload distribution. In contrast, RDD divides the simulation domain into equal-sized regular grid cells, which may not accurately reflect the fish's spatial distribution and can lead to subdomains with significantly different fish counts, resulting in load imbalance.

In summary, our results clearly demonstrate that the Voronoi-based domain decomposition approach proves to be a more effective and efficient method for achieving load balance in simulations of fish behavior. This improvement holds the potential to yield significant enhancements in computational performance.

\begin{figure}[H]
    \centering
    \subfloat[\centering Regular Cells  operation]{{\includegraphics[scale=0.5]{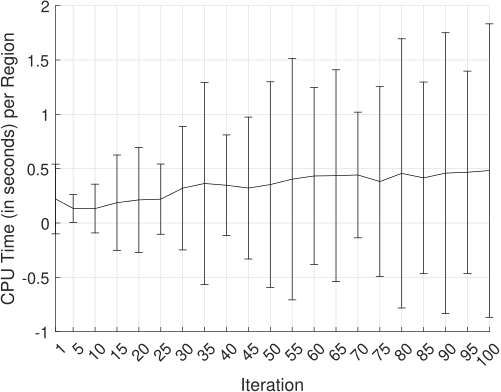} }}%
    \qquad
    \subfloat[\centering Voronoi Cells operations]{{\includegraphics[scale=0.5]{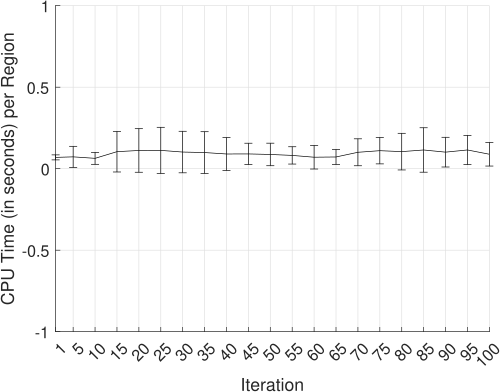} }}%
    \caption{The graph shows the mean (represented by the curve) and standard deviation (indicated by vertical lines) of the computational cost per region per iteration. This simulation involves 5,000 agents.}%
    \label{fig:CPU_regions}%
\end{figure}

Likewise, the graphs in Fig. \ref{fig:CPU_regions} illustrate the superior load balancing performance of our approach compared to the regular grid cells (RDD) implementation within each subdomain of the simulation. In comparison to the RDD approach (as shown in Fig. \ref{fig:CPU_regions}a), our approach exhibits significantly reduced variation in the computational cost of simulation within each subdomain, aligning more closely with the average workload (as seen in Fig. \ref{fig:CPU_regions}b).

\section{Conclusions and Future Work}
\label{con}
We have successfully implemented a large-scale simulation of fish schooling behavior using Voronoi tessellations generated by fuzzy c-Mean cluster centroids. Our approach effectively addresses common challenges in agent-based simulations, including communication overload and workload imbalances. Our simulations have shown that our partitioning algorithm maintains spatial and temporal load balance while significantly reducing computational costs compared to other methods.

Looking ahead, our future work will involve applying this approach to simulate fish migratory behaviors, incorporating environmental variables like temperature, salinity, and ocean currents. Given the varying resolutions of these variables, this simulation will be more complex, necessitating efficient domain decomposition techniques to mitigate communication overload and high computational costs.
\printbibliography

@inproceedings{reynolds1987flocks,
  title={Flocks, herds and schools: A distributed behavioral model},
  author={Reynolds, Craig W},
  booktitle={Proceedings of the 14th annual conference on Computer graphics and interactive techniques},
  pages={25--34},
  year={1987}
}

@article{barbaro2009discrete,
  title={Discrete and continuous models of the dynamics of pelagic fish: application to the capelin},
  author={Barbaro, Alethea BT and Taylor, Kirk and Trethewey, Peterson F and Youseff, Lamia and Birnir, Bj{\"o}rn},
  journal={Mathematics and Computers in Simulation},
  volume={79},
  number={12},
  pages={3397--3414},
  year={2009},
  publisher={Elsevier}
}

@inproceedings{chen2006genetic,
  title={Genetic algorithms for optimization of boids model},
  author={Chen, Yen-Wei and Kobayashi, Kanami and Huang, Xinyin and Nakao, Zensho},
  booktitle={International Conference on Knowledge-Based and Intelligent Information and Engineering Systems},
  pages={55--62},
  year={2006},
  organization={Springer}
}

@inproceedings{alaliyat2014optimisation,
  title={Optimisation Of Boids Swarm Model Based On Genetic Algorithm And Particle Swarm Optimisation Algorithm (Comparative Study).},
  author={Alaliyat, Saleh and Yndestad, Harald and Sanfilippo, Filippo},
  booktitle={ECMS},
  pages={643--650},
  year={2014},
  organization={Citeseer}
}

@misc{barksten2013extending,
  title={Extending Reynolds’ flocking model to asimulation of sheep in the presence of a predator},
  author={Barksten, Martin and Rydberg, David},
  year={2013}
}

@misc{chang2019investigating,
  title={Investigating and Modeling the Emergent Flocking Behaviour of Sheep Under Threat with Fear Contagion},
  author={Chang, Gabriel and Stjerndal, Michaela},
  year={2019}
}

@article{hartman2006autonomous,
  title={Autonomous boids},
  author={Hartman, Christopher and Benes, Bedrich},
  journal={Computer Animation and Virtual Worlds},
  volume={17},
  number={3-4},
  pages={199--206},
  year={2006},
  publisher={Wiley Online Library}
}

@article{tran2020switching,
  title={Switching formation strategy with the directed dynamic topology for collision avoidance of a multi-robot system in uncertain environments},
  author={Tran, Vu Phi and Garratt, Matthew A and Petersen, Ian R},
  journal={IET Control Theory \& Applications},
  volume={14},
  number={18},
  pages={2948--2959},
  year={2020},
  publisher={IET}
}

@article{erra2004massive,
  title={Massive simulation using gpu of a distributed behavioral model of a flock with obstacle avoidance},
  author={Erra, Ugo and De Chiara, Rosario and Scarano, Vittorio and Tatafiore, Maurizio},
  journal={Proceedings of Vision, Modeling and Visualization 2004 (VMV)},
  year={2004}
}

@article{ballerini2008interaction,
  title={Interaction ruling animal collective behavior depends on topological rather than metric distance: Evidence from a field study},
  author={Ballerini, Michele and Cabibbo, Nicola and Candelier, Raphael and Cavagna, Andrea and Cisbani, Evaristo and Giardina, Irene and Lecomte, Vivien and Orlandi, Alberto and Parisi, Giorgio and Procaccini, Andrea and others},
  journal={Proceedings of the national academy of sciences},
  volume={105},
  number={4},
  pages={1232--1237},
  year={2008},
  publisher={National Acad Sciences}
}

@inproceedings{dmytruk2021safe,
  title={Safe Tightly-Constrained UAV Swarming in GNSS-denied Environments},
  author={Dmytruk, Andriy and Nascimento, Tiago and Ahmad, Afzal and B{\'a}{\v{c}}a, Tom{\'a}{\v{s}} and Saska, Martin},
  booktitle={2021 International Conference on Unmanned Aircraft Systems (ICUAS)},
  pages={1391--1399},
  year={2021},
  organization={IEEE}
}

@inproceedings{subbey2003strategy,
  title={A strategy for rapid quantification of uncertainty in reservoir performance prediction},
  author={Subbey, Sam and Mike, Christie and Sambridge, Malcolm},
  booktitle={SPE Reservoir Simulation Symposium},
  year={2003},
  organization={OnePetro}
}

@inproceedings{na2010research,
  title={Research on k-means clustering algorithm: An improved k-means clustering algorithm},
  author={Na, Shi and Xumin, Liu and Yong, Guan},
  booktitle={2010 Third International Symposium on intelligent information technology and security informatics},
  pages={63--67},
  year={2010},
  organization={Ieee}
}

@article{sun2008clustering,
  title={Clustering algorithms research},
  author={Sun, Ji-Gui and Liu, Jie and Zhao, Lian-Yu},
  journal={Journal of software},
  volume={19},
  number={1},
  pages={48--61},
  year={2008}
}

@article{pokojski2018voronoi,
  title={Voronoi diagrams--inventor, method, applications},
  author={Pokojski, Wojciech and Pokojska, Paulina},
  journal={Polish Cartographical Review},
  volume={50},
  number={3},
  pages={141--150},
  year={2018}
}

@article{boots1999spatial,
  title={Spatial tessellations},
  author={Boots, B and Okabe, A and Sugihara, K},
  journal={Geographical information systems},
  volume={1},
  pages={503--526},
  year={1999},
  publisher={John Wiley \& Sons New York, NY}
}

@article{alglave2014herding,
  title={Herding cats: Modelling, simulation, testing, and data mining for weak memory},
  author={Alglave, Jade and Maranget, Luc and Tautschnig, Michael},
  journal={ACM Transactions on Programming Languages and Systems (TOPLAS)},
  volume={36},
  number={2},
  pages={1--74},
  year={2014},
  publisher={ACM New York, NY, USA}
}

@article{erneholm2011simulation,
  title={Simulation of the flocking behavior of birds with the boids algorithm},
  author={Erneholm, CARL-OSCAR},
  journal={Royal Institute of Technology},
  year={2011}
}

@article{huth1992simulation,
  title={The simulation of the movement of fish schools},
  author={Huth, Andreas and Wissel, Christian},
  journal={Journal of theoretical biology},
  volume={156},
  number={3},
  pages={365--385},
  year={1992},
  publisher={Elsevier}
}

@article{maire2015molecular,
  title={Molecular-level tuning of cellular autonomy controls the collective behaviors of cell populations},
  author={Maire, Th{\'e}o and Youk, Hyun},
  journal={Cell systems},
  volume={1},
  number={5},
  pages={349--360},
  year={2015},
  publisher={Elsevier}
}

@article{schranz2020swarm,
  title={Swarm robotic behaviors and current applications},
  author={Schranz, Melanie and Umlauft, Martina and Sende, Micha and Elmenreich, Wilfried},
  journal={Frontiers in Robotics and AI},
  volume={7},
  pages={36},
  year={2020},
  publisher={Frontiers Media SA}
}

@article{warren2018collective,
  title={Collective motion in human crowds},
  author={Warren, William H},
  journal={Current directions in psychological science},
  volume={27},
  number={4},
  pages={232--240},
  year={2018},
  publisher={SAGE Publications Sage CA: Los Angeles, CA}
}

@article{treado2021bridging,
  title={Bridging particle deformability and collective response in soft solids},
  author={Treado, John D and Wang, Dong and Boromand, Arman and Murrell, Michael P and Shattuck, Mark D and O'Hern, Corey S},
  journal={Physical Review Materials},
  volume={5},
  number={5},
  pages={055605},
  year={2021},
  publisher={APS}
}

@article{egan2013design,
  title={Design of complex biologically based nanoscale systems using multi-agent simulations and structure--behavior--function representations},
  author={Egan, Paul F and Cagan, Jonathan and Schunn, Christian and LeDuc, Philip R},
  journal={Journal of Mechanical Design},
  volume={135},
  number={6},
  pages={061005},
  year={2013},
  publisher={American Society of Mechanical Engineers}
}

@article{solar2011high,
  title={High performance distributed cluster-based individual-oriented fish school simulation},
  author={Solar, Roberto and Suppi, Remo and Luque, Emilio},
  journal={Procedia Computer Science},
  volume={4},
  pages={76--85},
  year={2011},
  publisher={Elsevier}
}

@article{vigueras2010comparative,
  title={A comparative study of partitioning methods for crowd simulations},
  author={Vigueras, Guillermo and Lozano, Miguel and Orduna, Juan Manuel and Grimaldo, Francisco},
  journal={Applied Soft Computing},
  volume={10},
  number={1},
  pages={225--235},
  year={2010},
  publisher={Elsevier}
}

@article{solar2012proximity,
  title={Proximity load balancing for distributed cluster-based individual-oriented fish school simulations},
  author={Solar, Roberto and Suppi, Remo and Luque, Emilio},
  journal={Procedia Computer Science},
  volume={9},
  pages={328--337},
  year={2012},
  publisher={Elsevier}
}

@article{stijnman2003partitioning,
  title={Partitioning 3D space for parallel many-particle simulations},
  author={Stijnman, MA and Bisseling, RH and Barkema, GT},
  journal={Computer physics communications},
  volume={149},
  number={3},
  pages={121--134},
  year={2003},
  publisher={Elsevier}
}

@article{zhang2009fast,
  title={A fast adaptive load balancing method for parallel particle-based simulations},
  author={Zhang, Dongliang and Jiang, Changjun and Li, Shu},
  journal={Simulation Modelling Practice and Theory},
  volume={17},
  number={6},
  pages={1032--1042},
  year={2009},
  publisher={Elsevier}
}

@article{egorova2019parallel,
  title={Parallel SPH modeling using dynamic domain decomposition and load balancing displacement of Voronoi subdomains},
  author={Egorova, Maria Serg and Dyachkov, Sergey A and Parshikov, Anatoly N and Zhakhovsky, VV},
  journal={Computer Physics Communications},
  volume={234},
  pages={112--125},
  year={2019},
  publisher={Elsevier}
}

@inproceedings{youseff2008parallel,
  title={Parallel modeling of fish interaction},
  author={Youseff, Lamia and Barbaro, Alethea and Trethewey, Peterson and Birnir, Bj{\"o}rn and Gilbert, John R},
  booktitle={2008 11th IEEE International Conference on Computational Science and Engineering},
  pages={234--241},
  year={2008},
  organization={IEEE}
}

@inproceedings{zhou2004parallel,
  title={Parallel simulation of group behaviors},
  author={Zhou, Bo and Zhou, Suiping},
  booktitle={Proceedings of the 2004 Winter Simulation Conference, 2004.},
  volume={1},
  year={2004},
  organization={IEEE}
}

@article{cordasco2018distributed,
  title={Distributed mason: A scalable distributed multi-agent simulation environment},
  author={Cordasco, Gennaro and Scarano, Vittorio and Spagnuolo, Carmine},
  journal={Simulation Modelling Practice and Theory},
  volume={89},
  pages={15--34},
  year={2018},
  publisher={Elsevier}
}

@article{solar2010high,
  title={High performance individual-oriented simulation using complex models},
  author={Solar, Roberto and Suppi, Remo and Luque, Emilio},
  journal={Procedia Computer Science},
  volume={1},
  number={1},
  pages={447--456},
  year={2010},
  publisher={Elsevier}
}

@inproceedings{wang2009cluster,
  title={Cluster based partitioning for agent-based crowd simulations},
  author={Wang, Yongwei and Lees, Michael and Cai, Wentong and Zhou, Suiping and Low, Malcolm Yoke Hean},
  booktitle={Proceedings of the 2009 Winter Simulation Conference (WSC)},
  pages={1047--1058},
  year={2009},
  organization={IEEE}
}

@inproceedings{labba2015towards,
  title={Towards a conceptual framework to support adaptative agent-based systems partitioning},
  author={Labba, Chahrazed and ben Saoud, Narjes Bellamine and Dugdale, Julie},
  booktitle={2015 IEEE/ACIS 16th International Conference on Software Engineering, Artificial Intelligence, Networking and Parallel/Distributed Computing (SNPD)},
  pages={1--5},
  year={2015},
  organization={IEEE}
}

@article{marzouk2005k,
  title={K-means clustering for optimal partitioning and dynamic load balancing of parallel hierarchical N-body simulations},
  author={Marzouk, Youssef M and Ghoniem, Ahmed F},
  journal={Journal of Computational Physics},
  volume={207},
  number={2},
  pages={493--528},
  year={2005},
  publisher={Elsevier}
}

@article{araki2022dynamic,
  title={Dynamic load balancing with over decomposition in plasma plume simulations},
  author={Araki, Samuel J and Martin, Robert S},
  journal={Journal of Parallel and Distributed Computing},
  volume={163},
  pages={136--146},
  year={2022},
  publisher={Elsevier}
}

@inproceedings{von2018balanced,
  title={Balanced k-means for parallel geometric partitioning},
  author={von Looz, Moritz and Tzovas, Charilaos and Meyerhenke, Henning},
  booktitle={Proceedings of the 47th International Conference on Parallel Processing},
  pages={1--10},
  year={2018}
}

@article{wu2021k,
  title={k-Means Clustering Algorithm and Its Simulation Based on Distributed Computing Platform},
  author={Wu, Chunqiong and Yan, Bingwen and Yu, Rongrui and Yu, Baoqin and Zhou, Xiukao and Yu, Yanliang and Chen, Na},
  journal={Complexity},
  volume={2021},
  year={2021},
  publisher={Hindawi}
}

@article{zhu2020improved,
  title={Improved soft-k-means clustering algorithm for balancing energy consumption in wireless sensor networks},
  author={Zhu, Botao and Bedeer, Ebrahim and Nguyen, Ha H and Barton, Robert and Henry, Jerome},
  journal={IEEE Internet of Things Journal},
  volume={8},
  number={6},
  pages={4868--4881},
  year={2020},
  publisher={IEEE}
}

@article{hanebutte1995traffic,
  title={Traffic simulations on parallel computers using domain decomposition techniques},
  author={Hanebutte, UR and Tentner, Adrian Michel and others},
  year={1995}
}

@article{shen2014distributed,
  title={Distributed information theoretic clustering},
  author={Shen, Pengcheng and Li, Chunguang},
  journal={IEEE Transactions on Signal Processing},
  volume={62},
  number={13},
  pages={3442--3453},
  year={2014},
  publisher={IEEE}
}

@article{yang2019resource,
  title={Resource assignment based on dynamic fuzzy clustering in elastic optical networks with multi-core fibers},
  author={Yang, Hui and Yao, Qiuyan and Yu, Ao and Lee, Young and Zhang, Jie},
  journal={IEEE Transactions on Communications},
  volume={67},
  number={5},
  pages={3457--3469},
  year={2019},
  publisher={IEEE}
}

@book{bezdek2013pattern,
  title={Pattern recognition with fuzzy objective function algorithms},
  author={Bezdek, James C},
  year={2013},
  publisher={Springer Science \& Business Media}
}
\end{document}